\documentclass[aps,prl,twocolumn,notitlepage,superscriptaddress]{revtex4-1}
\usepackage[colorlinks=true, citecolor=magenta, linkcolor=blue, urlcolor=blue]{hyperref}
\usepackage{graphicx}
\usepackage{amsmath}
\usepackage{amssymb}
\usepackage{amsfonts}
\usepackage{hyperref}
\usepackage{mathtools}
\usepackage{xcolor}
\usepackage{tikz}
\usepackage{enumitem}
\usetikzlibrary{shapes.geometric, arrows}
\usetikzlibrary{decorations.markings}

\usepackage[T1]{fontenc}
\usepackage{textcomp, amssymb}
\usepackage[utf8]{inputenc}

\allowdisplaybreaks

\begin{document}
 
\title{Comparative Analysis of THz Signal Emission from SiO$_2$/CoFeB/Metal Heterostructures: Wideband and High-Frequency THz Signal Advantage of PtBi-based Emitter}

\author{Tristan Joachim Winkel}
\email{tristan.winkel@uni-greifswald.de}
\affiliation{Institut für Physik, Universität Greifswald, Greifswald, Germany}
\author{Tahereh Sadat Parvini}
\email{tahereh.parvini@uni-greifswald.de}
\affiliation{Institut für Physik, Universität Greifswald, Greifswald, Germany}
\author{Finn-Frederik Stiewe}
\affiliation{Institut für Physik, Universität Greifswald, Greifswald, Germany}
\author{Jakob Walowski}
\affiliation{Institut für Physik, Universität Greifswald, Greifswald, Germany}
\author{Farshad Moradi}
\affiliation{ICELab, Aarhus University, Denmark}
\author{Markus M\"unzenberg}
\affiliation{Institut für Physik, Universität Greifswald, Greifswald, Germany}
\date{\today}

\begin{abstract}
Spintronic THz emitters have attracted much attention due to their desirable properties, such as affordability, ultra-wideband capability, high efficiency, and tunable polarization. In this study, we investigate the characteristics of THz signals, including their frequency, bandwidth, and amplitude, emitted from a series of heterostructures with ferromagnetic (FM) and nonmagnetic (NM) materials. The FM layer consists of a wedge-shaped CoFeB layer with a thickness of 0 to 5 nm, while the NM materials include various metals such as Pt, Au, W, Ru, Pt$_{\%92}$Bi$_{\%8}$, and Ag$_{\%90}$Bi$_{\%10}$ alloys. Our experiments show that the emitter with Pt-NM layer has the highest amplitude of the emitted THz signal. However, the PtBi-based emitter exhibits a higher central THz peak and wider bandwidth, making it a promising candidate for broadband THz emitters. These results pave the way for further exploration of the specific compositions of Pt$_{1-x}$Bi$_{x}$ for THz emitter design, especially with the goal of generating higher frequency and wider bandwidth THz signals. These advances hold significant potential for applications in various fields such as high-resolution imaging, spectroscopy, communications, medical diagnostics, and more.

\end{abstract}

\maketitle

\section{Introduction}

The Hall effect, discovered in 1879 \cite{hall1880new}, is a crucial technique in modern measurement technology, allowing for the determination of magnetic field strength and charge carrier type, density, and mobility. The spin Hall effect (SHE), which is the quantum mechanical analogue of the 
Magnus effect, was predicted in 1971 \cite{d1971possibility} and later experimentally confirmed in the early 2000s \cite{murakami2003dissipationless, kato2004observation, sinova2004universal}. The SHE arises from the interplay between the electron's spin and motion in the presence of spin-orbit coupling, and has significant implications for spintronics and related technologies such as the generation of terahertz (THz) radiation, high-speed data processing, and quantum information processing \cite{demidov2014nanoconstriction, sinova2015spin, jungwirth2012spin}. THz waves have potential applications in materials science, biology, and medicine, and further research could broaden their use in spectroscopy, imaging, and communication \cite{tonouchi2007cutting, cheon2019detection, pawar2013terahertz, gowen2012terahertz, henriksen2022plastic, wang2014ultrabroadband, seifert2016efficient}. Typical THz emitters are made up of ferromagnetic (FM)/nonmagnetic (NM) heterostructures where the FM layer is pumped using a femtosecond laser to generate a spin-dependent excitation of electrons. The inverse spin Hall effect (ISHE) in the NM layer converts the longitudinal spin-polarized current ($\mathbf{J}_{s}$) into a transient transverse current ($\mathbf{J}_{c}$) according to $\mathbf{J}_{c}=\theta_{SH}\mathbf{J}_{s}\times\frac{\mathbf{M}}{|\mathbf{M}|}$, resulting in the emission of terahertz radiation \cite{kampfrath2013terahertz, battiato2010superdiffusive, zhukov2006lifetimes, gradhand2010extrinsic}. $\theta_{SH}$ is the spin Hall angle that characterizes the efficiency of the spin Hall effect and can be enhanced through various means such as selecting materials with large spin-orbit coupling, optimizing material properties including crystal structure, impurity level, and thickness, as well as controlling experimental conditions. By precisely determining the materials and thickness in THz emitters, it becomes possible to optimize the amplitude, frequency, and bandwidth of the THz field.

In this study, we fabricated a series of FM/NM heterostructures to conduct a comparative study on the characteristics of the emitted THz signal. The FM layer consists of a wedge layer of Co$_{40}$Fe$_{40}$B$_{20}$ with a thickness ranging from 0 to 5 nm. Meanwhile, the NM layer comprised a diverse range of materials, including Pt (2, 3, and 4 nm), W (2 nm), Au (2 nm), Ru (4 nm), as well as the alloys Pt$_{\%92}$Bi$_{\%8}$ (2 nm) and Ag$_{\%90}$Bi$_{\%10}$ (2 nm). Our investigation focused on evaluating the amplitude, central frequency, and bandwidth of the THz signal emitted from all samples. Furthermore, we analyzed the determining factors that influenced our observations.

\section{Materials and Methods}

In this study, we adopted a conventional approach of employing the substrate/FM/NM layer configuration for the fabrication of THz emitters. Initially, a magnetic layer consisting of Co$_{40}$Fe$_{40}$B$_{20}$ (CoFeB) was deposited onto a fused silica substrate using magnetron-sputtering technique. To explore the influence of FM-layer thickness on the characteristics of the emitted THz signal, we designed the layer in a wedge shape, allowing for a variable thickness ranging from 0 to 5 nm. To investigate the contribution of the NM-layer material to the terahertz signal, we examined various pure heavy metals such as Pt, W, Au, Ru , as well as alloys Pt$_{\%92}$Bi$_{\%8}$ and Ag$_{\%90}$Bi$_{\%10}$ as NM-layers. Furthermore, We fabricated a series of heterostructures with varying thicknesses of pure Pt layers to study the impact of Pt layer thickness on the emitted THz signal. The NM-layers were deposited on the CoFeB using electron beam (e-beam) evaporation. The alloys are formed through a controlled combined deposition of both constituent materials. The ratio between the alloyed materials is determined by adjusting the deposition rates, monitored using quartz crystal monitoring. The required ratio of the deposition rates ($v_a/v_b$) can be calculated using the following formula
\begin{equation}
    \frac{v_a}{v_b}= \frac{M_a\times\rho_b}{M_b\times\rho_a}.r,
\end{equation}
where $\rho$ represents the density, $M$ denotes the molar mass, and $r$ signifies the desired molar ratio of the alloyed materials. Due to the high sensitivity of spin currents to interface contaminations or oxidation, we implemented strict precautions during sample fabrication. This involved conducting the fabrication process and transferring the samples between chambers under high vacuum conditions, ensuring the preservation of interface integrity without any modifications.

\section{Results and Discussion}

\begin{figure}[t!]
  \centering
  \includegraphics[width=0.4\textwidth]{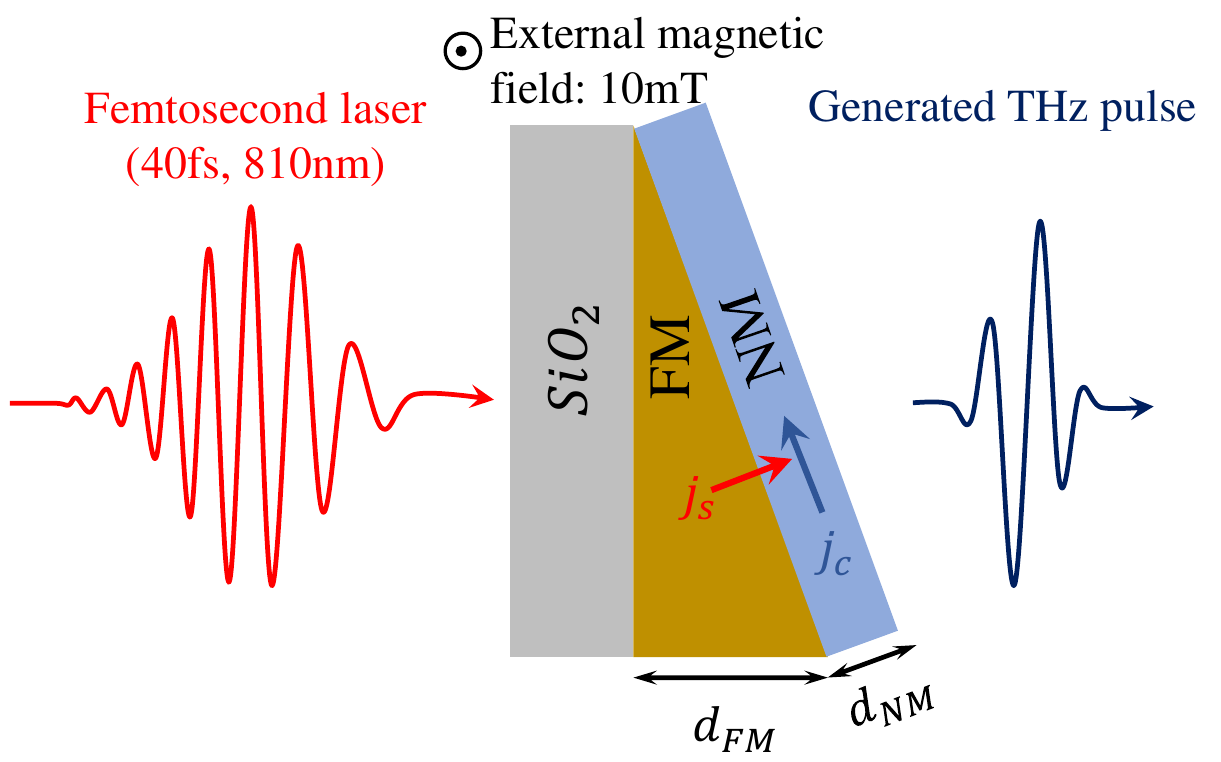} 
  \caption{Side view of a FM/NN bilayer consisting of a ferro-magnetic metal layer (FM) and an adjacent nonmagnetic metal layer (NM). A femtosecond laser pulse excites the metal stack from the substrate side.}
\label{fig1}
\end{figure}

Our previous publication \cite{stiewe2022spintronic} have provided a detailed description of our experimental setup for generating and detecting THz signals from spintronic emitters. Fig. \ref{fig1} provides a simple schematic of our THz emitters and the corresponding mechanism for THz generation. We generate THz radiation in heterostructures by employing femtosecond laser pulses with central wavelengths of $\sim$810 nm, pulse durations of 40 fs, and repetition rates of 80 MHz. The detection of THz signals is carried out using commercial low-temperature grown-GaAs (LT-GaAs) Auston switches with a bandwidth greater than 4 THz. To obtain a temporal THz spectrum at each position of the sample, we utilized a two dimensional scanning technique with motorized stages. A comprehensive description of the methodology can be found in \cite{stiewe2022magnetic}. Additionally, the samples were subjected to scanning using a magneto-optic Kerr effect setup to investigate the relationship between the position and the thickness of the CoFeB layer. The obtained data was fitted using an error function and combined with the data acquired from the THz setup. This integration allowed us to determine the THz emission corresponding to each sample as a function of the CoFeB layer thickness.

\begin{figure*}[ht]
  \centering
  \includegraphics[width=0.32\textwidth]{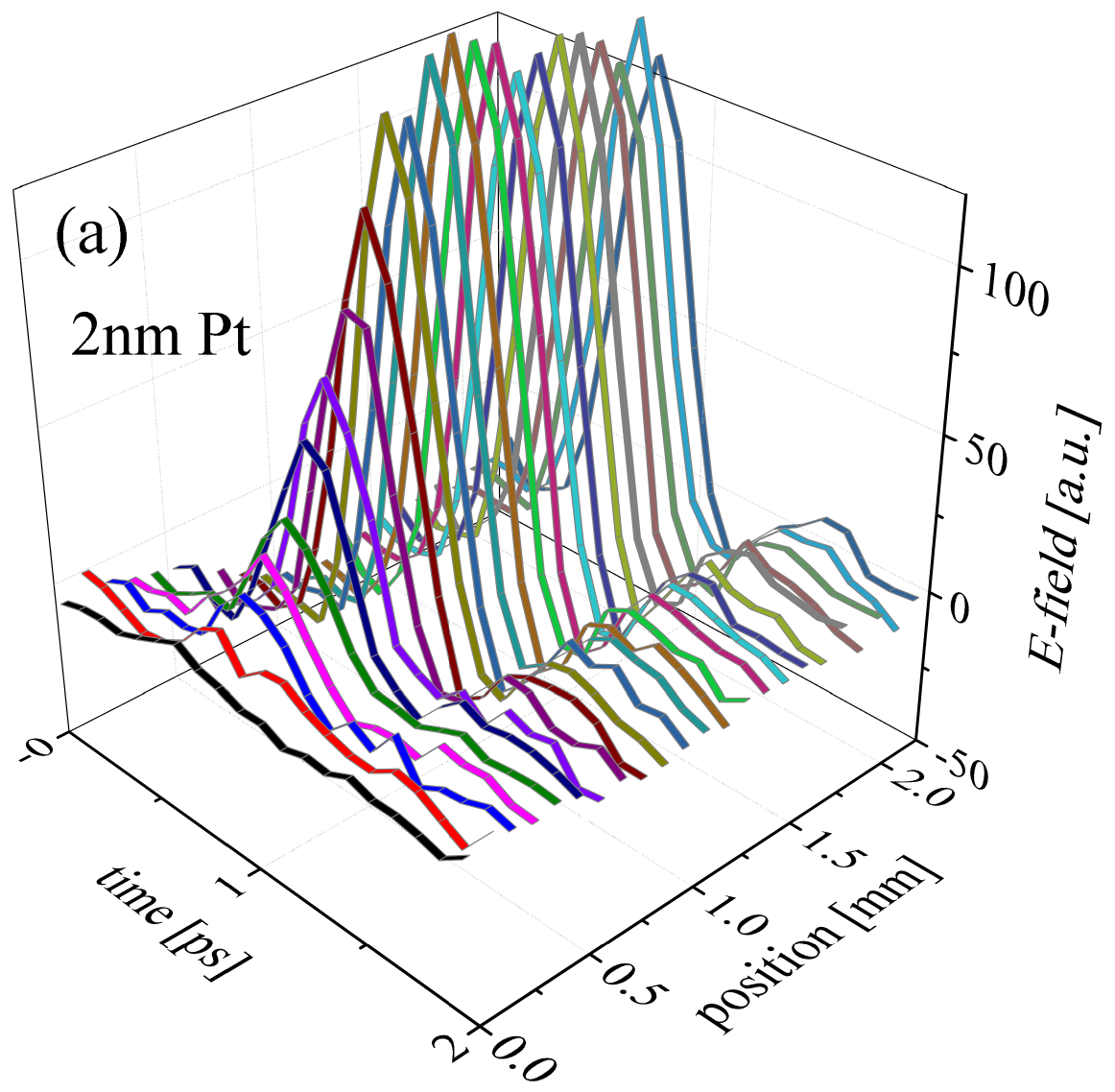} 
  \includegraphics[width=0.32\textwidth]{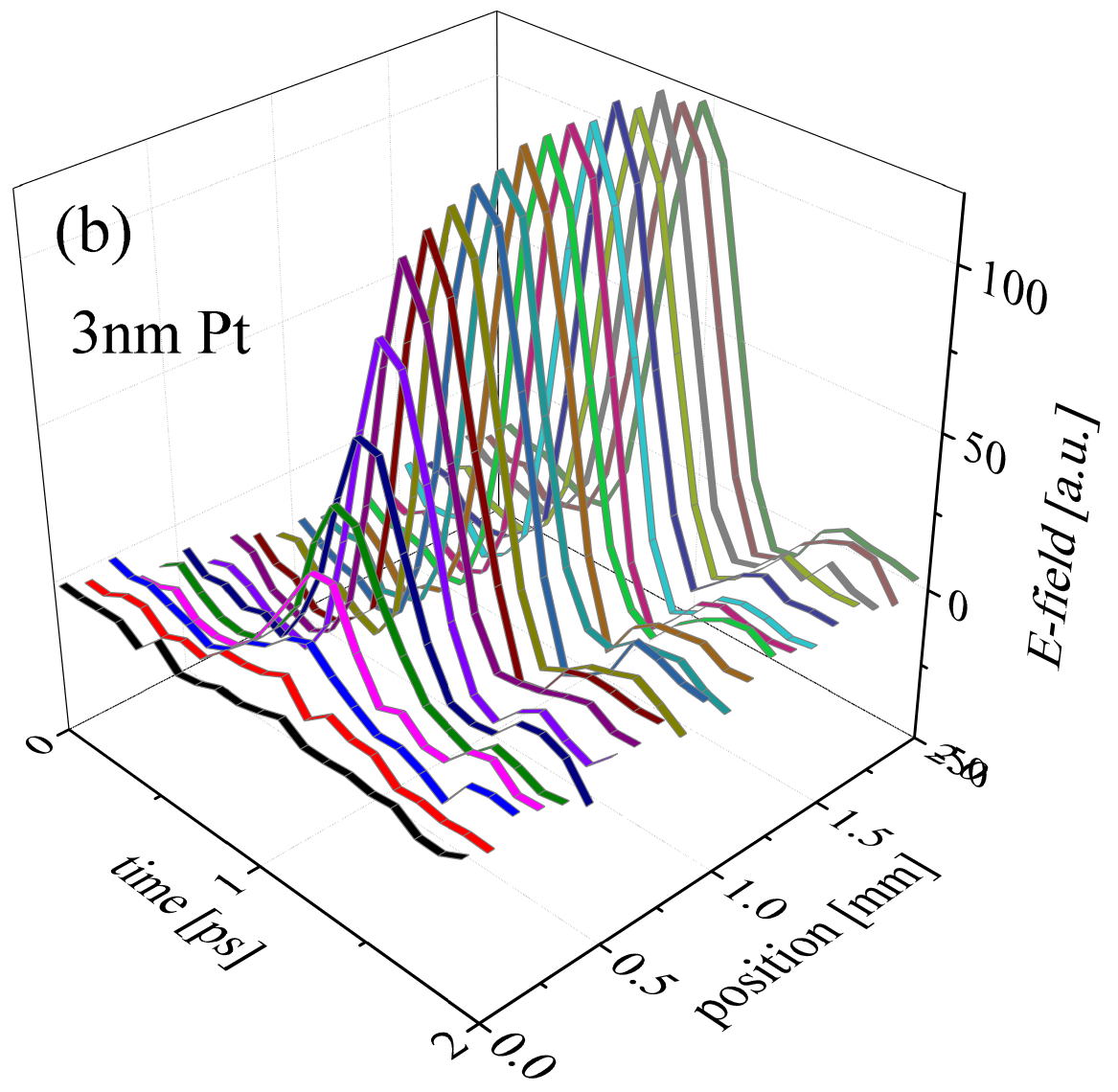}\\
  \includegraphics[width=0.32\textwidth]{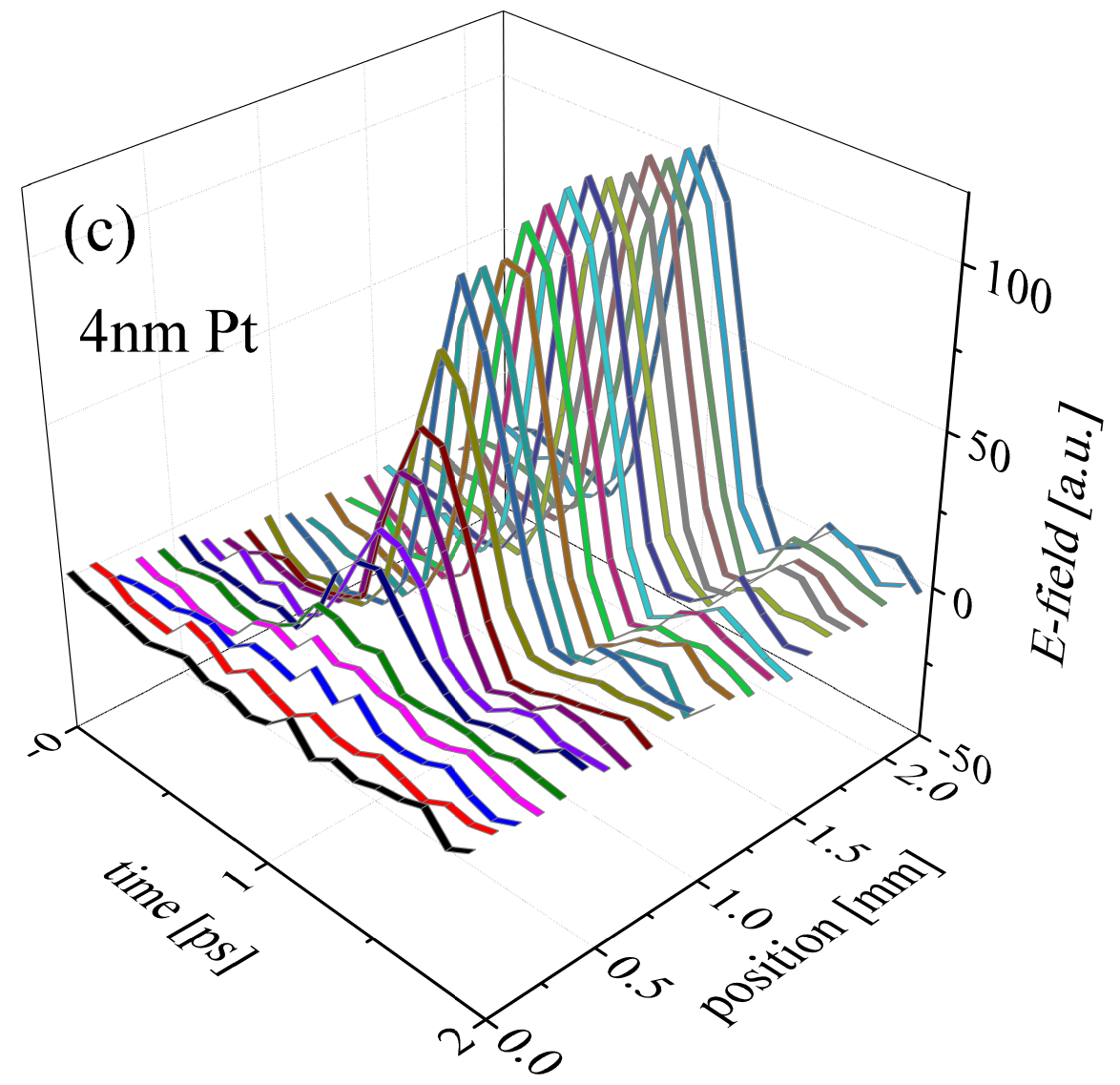} 
  \includegraphics[width=0.32\textwidth]{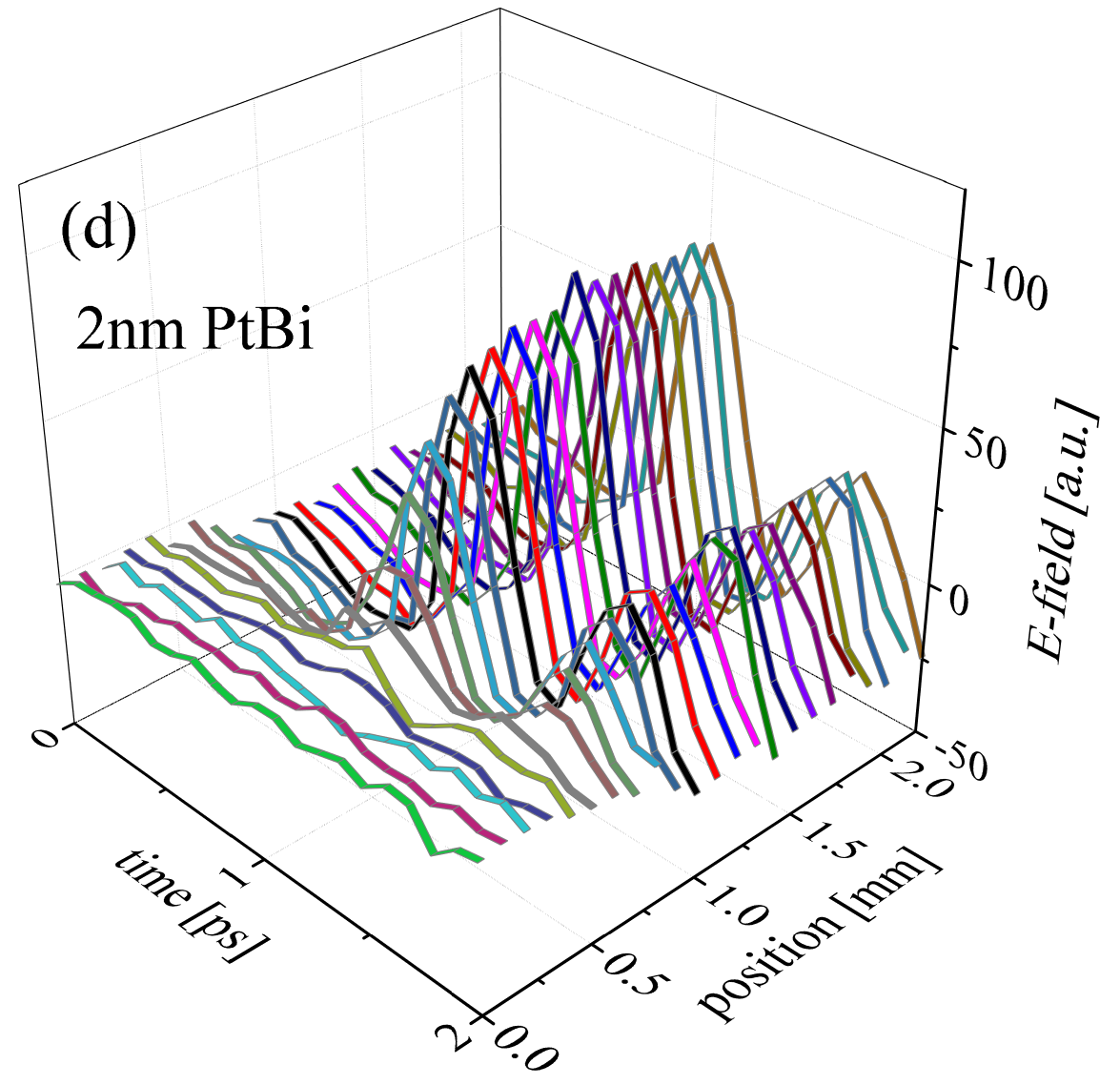}
  \caption{THz signal amplitude as a function of CoFeB layer thickness and pump-probe delay time for stacks with Pt and PtBi NM layer. (a) 2nm Pt, (b) 3nm Pt, (c) 4nm Pt, and (d) 2nm PtBi alloy. The pump fluence is 0.8$\pm$0.2$mJ/cm^2$.}
\label{fig2}
\end{figure*}
 Under the illumination of ultrafast femtosecond (fs) laser pulses, spin-polarized electrons ($j_s$) within the FM layer get excited and diffuse into the NM layer, forming a spin current. The inverse spin Hall effect (ISHE) comes into play, converting this spin current into a transient transverse charge current ($j_c$) within the NM layer, ultimately leading to the emission of terahertz (THz) radiation. In the plane-wave approximation, the amplitude of the electric field of the emitted THz signal reads \cite{seifert2022spintronic}
\begin{equation}
E_{THz}=\frac{AF}{d_{NM}+d_{FM}}.j_{s}^{0}t_{\frac{FM}{NM}}\lambda_{NM}tanh\frac{d_{NM}}{\lambda_{NM}}.\theta_{SH}.eZ(\omega),
\label{eq2}
\end{equation}
where the terms respectively represent pump-pulse absorption, spin-current generation, spin-to-charge current conversion, and charge-current-to-electric-field conversion. The parameters A, $j_{s}^{0}$, $t_{\frac{FM}{NM}}$, $\lambda_{NM}$, and $\theta_{SH}$ represent the absorbed fraction of the incident pump-pulse fluence ($F$), the generated spin-current density per pump-pulse excitation density, the interfacial spin-current transmission amplitude between the FM and NM layers, the spin-current relaxation length in the NM layer, and the spin Hall angle specific to the NM material, respectively. The frequency-dependent impedance of the emitter, denoted as $Z(\omega)$, is given by $Z(\omega)=\frac{Z_{0}}{n_{1}+n_{2}+Z_{0}G}$, where $Z_0\approx377\Omega$ represents the free-space impedance, and $n_1(\omega)$ and $n_2(\omega)$ are the refractive indices of air and the substrate, respectively, and $G(\omega)$ represents the THz sheet conductance. Consequently, by meticulously controlling the parameters in Eq. \ref{eq2}, it is possible to achieve the desired signal. Subsequently, we proceed to investigate the role of these parameters in the emitted THz signal from our fabricated emitters.
\\
{\bf{Dependence of THz Signal Amplitude, Central Frequency, and Bandwidth on Heterostructure Thickness and NM Materials:}} Fig. \ref{fig2} illustrates the emitted THz signal as a function of delay time and laser spot position for distinct samples. Accordingly, irrespective of the type of metal layers, no THz emission is detected at the thin end of the wedge, where the CoFeB thickness is 0 nm. As the thickness of the CoFeB layer increases, the amplitude of the emitted THz signal exhibits growth, peaking at around 2 nm of CoFeB thickness. It is noteworthy that beyond this critical point, further increases in CoFeB thickness do not have a significant impact on the amplitude of the THz signal. This behavior can be attributed to the progressive diffusion of an increasing number of electrons into the NM layer as the CoFeB layer thickness increases up to 2 nm. The influx of electrons enhances the induced spin current, thereby generating stronger THz radiation. However, beyond a 2 nm thickness of the CoFeB layer, the presence of heightened structural defects, electron scattering, and resistance within the magnetic layer hinders further amplification of the THz signal amplitude. As a result, the THz signal reaches a plateau beyond this critical thickness. Furthermore, the results indicate that the maximum THz amplitude is attained when utilizing a 2 nm Pt layer as the heavy metal component of the emitter. The decrease in THz amplitude with increasing Pt thickness can be attributed to the absorption and attenuation of THz waves by the Pt layer, which will be discussed in the subsequent section. 

\begin{figure*}[ht]
  \centering
  \includegraphics[width=0.76\textwidth]{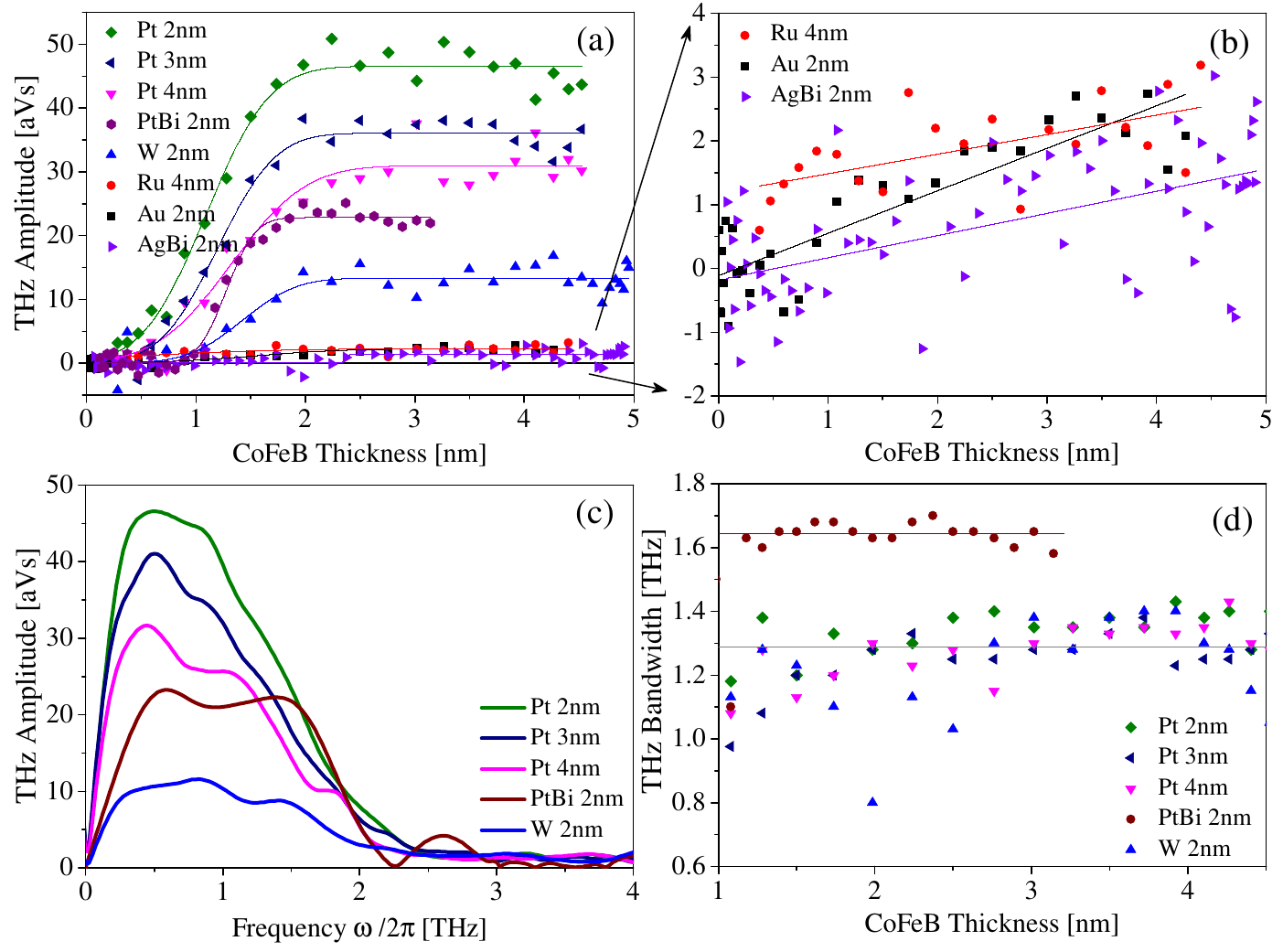} 
  \caption{(a) The amplitude of the THz signal at a frequency of 1 THz as a function of CoFeB-layer thicknesses for stacks containing different NM layers and varying thicknesses. The amplitude is measured in attovolt-seconds. (b) A magnified scale of the THz signal for stacks comprising NM layers Ru, Au, and AgBi. (c) The amplitude of the emitted THz signal as a function of frequency for emitters with a constant CoFeB thickness of 2 nm. (d) The THz bandwidth as a function of CoFeB-layer thicknesses. }
\label{fig3}
\end{figure*}

\begin{table*}[ht]
    \caption{Saturated THz amplitude and central frequency of the THz pulse emitted from heterostructures CoFeB (2nm)/NM.}
    \centering
    \begin{tabular}{l c c c c c c c c c c c c c c c c}
    \hline \hline
     NM Layer  &&  2nm Pt  &&  3nm Pt &&  4nm Pt &&  2nm PtBi  && 2nm W   &&  4nm Ru  &&  2nm Au  && 2nm AgBi \\ [1ex]
    \hline
    Amplitude [aVs] &&  46.35   &&  36.02  &&  30.83  &&  22.90     && -13.27  &&  2.21    &&  2.15    &&  1.32 \\ [1ex]
    
    Frequency [THz] &&  0.75   &&  0.73  &&  0.71  &&  1.04  && 0.84  &&  ---   &&  ---    &&  --- \\
    \hline \hline
\label{table1}
\end{tabular}
\end{table*}

To gain deeper insights into the fabricated THz emitters, we analyzed the correlation between the amplitude of the emitted THz field at 1 THz and the varying thicknesses of the CoFeB layer in stacks with specified NM layers, see Fig. \ref{fig3} (a) and (b). The THz emitters composed of pure Pt layers demonstrate the highest THz amplitude among the emitters. Specifically, the stack with a 2 nm Pt layer exhibits a THz signal amplitude that is twice as high as the stack with a 2 nm PtBi layer. However, the emitter with PtBi, as shown in Figs. \ref{fig3} (c) and (d), exhibits a wider bandwidth of approximately 0.35 THz compared to other stacks, along with a higher central frequency of the THz signal. Interestingly, it demonstrates a significant central frequency shift of approximately 0.3 THz when compared to the emitter with 2 nm Pt. These findings highlight the crucial role of the NM layer material in influencing THz characteristics and suggest the potential advantages of the PtBi stack for applications requiring broader bandwidth and higher frequencies, despite a lower THz amplitude. When evaluating the trade-off between THz signal strength and bandwidth, it is crucial to consider the specific requirements of the intended application, as different applications may prioritize either a higher THz signal strength or a broader bandwidth based on their unique needs and constraints. This observation suggests potential advantages in applications that benefit from higher frequency THz signals, including high-resolution imaging, spectroscopy, communications, medical diagnostics, and etc. The values of the saturated THz amplitudes and THz peak position measured for different emitters are presented in Table \ref{table1}. 
\\
\begin{figure}[ht]
  \centering
  \includegraphics[width=0.47\textwidth]{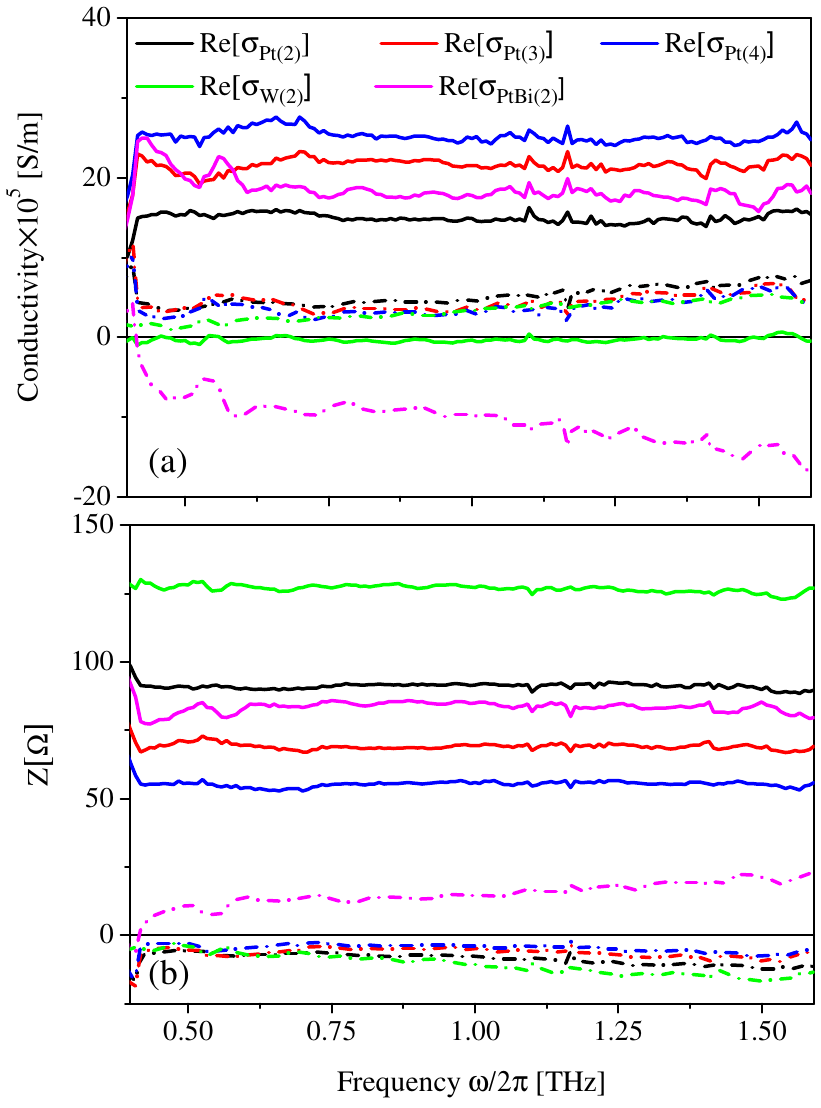} 
  \caption{Real (solid line) and imaginary (dashed-dotted line) parts of (a) conductivity, $\sigma(\omega)$, and (b) impedance, Z($\omega$), of different heavy metal thin films obtained via THz-TDS.}
\label{fig4}
\end{figure}
\\
{\bf{Exploring Conductance of NM Layers through Terahertz Time-Domain Spectroscopy (THz-TDS):}} The conductance of the NM layer, as indicated by Eq.\ref{eq2}, is a significant factor affecting the THz signal amplitude. To determine the conductance value of each NM layer, we direct the THz transient signal onto both the bare substrate and the substrate coated with a thin metal film. By measuring the THz transmission through the sample ($T_s(\omega)$) relative to that through the bare substrate ($T_{\text{sub}}(\omega)$), we can infer the conductance of the metal at THz frequencies using Tinkham’s formulas \cite{gruner1998millimeter, kamboj2017probing, seifert2018terahertz, tinkham1956energy}: 
\begin{equation}
    \frac{T_{s}(\omega)}{T_{sub}(\omega)}=\frac{E_{s}(\omega)}{E_{sub}(\omega)}=\frac{1+n_{sub}}{1+n_{sub}+Z_{0}G(\omega)},
\end{equation}
here, $E_{s}(\omega)$ and $E_{sub}(\omega)$ are the Fourier Transforms of the time-dependent THz electric field waveforms transmitted through the sample and substrate, respectively. The refractive index of the substrate is obtained using the formula $n_{\text{sub}}=\frac{c\Delta\phi}{d_{\text{sub}}\omega}+1$, where $\Delta\phi$ is the phase difference of the pulsed THz radiation incident upon the substrate ($E_i$) and transmitted through the substrate ($E_{\text{sub}}$). For a SiO$_2$ substrate with a thickness of $d_{\text{sub}}=500\mu$m, applying this method yields $n_{\text{sub}}\approx$1.98. 

Fig. \ref{fig4}(a) shows the real and imaginary parts of the optical conductivity for different NM thin films, obtained through the formula $G(\omega) = \sigma(\omega) \cdot d_{\text{NM}}$. The observed inequality $\left|\sigma_{Pt(2)}\right|<\left|\sigma_{Pt(3)}\right|<\left|\sigma_{Pt(4)}\right|$ demonstrates a positive correlation between film thickness and conductivity. Increasing the film thickness improves conductivity due to factors such as an increased number of charge carriers, reduced interfacial scattering, improved crystallinity, and decreased surface roughness. The Drude model \cite{jackson2021classical} was utilized to fit the conductivity of the Pt layers, enabling the determination of plasma frequency ($\omega_p=\sqrt{Ne^{2}/m\varepsilon_{0}}$, where N, e, m, $\varepsilon_{0}$ are  the carrier electron density, electron charge, electron effective mass, and vacuum permittivity, respectively) and scattering time ($\tau$). The findings demonstrate a clear trend of increasing $\omega_p$ with film thickness, indicating a proportional rise in carrier density ($\omega_{p,\text{Pt}(2)} = 0.19 \times 10^{16} \text{ Hz}$, $\omega_{p,\text{Pt}(3)} = 0.3 \times 10^{16} \text{ Hz}$, $\omega_{p,\text{Pt}(4)} = 0.36 \times 10^{16} \text{ Hz}$). Additionally, as the film thickness increases, the scattering time decreases ($\tau_{\text{Pt}(2)}=52f$s, $\tau_{\text{Pt}(3)}=32f$s, $\tau_{\text{Pt}(4)}=23f$s), resulting in a decrease in electron mobility ($\mu\propto\tau$). The impedance analysis in Fig. \ref{fig4}(b) confirms that thicker Pt layers exhibit lower impedance compared to the thinner layers, with a clear trend of $\left|Z_{Pt(2)}\right|>\left|Z_{Pt(3)}\right|>\left|Z_{Pt(4)}\right|$. This finding directly corresponds to a reduction in the emitted THz signal from thicker NM layer, as supported by Eq. \ref{eq2} and proved in Fig. \ref{fig2}. W exhibits greater impedance compared to other metals, but one contributing factor to its smaller THz amplitude is its relatively diminished spin Hall angle. Remarkably, PtBi (2nm) exhibits a real component of conductivity that is comparable to that of Pt (2nm). However, what sets it apart is its surpassing imaginary component, which not only exceeds that of Pt but also bears a negative value (Im$[\sigma_{PtBi}]=-a\omega$, where $a$ is a constant). This can not be explained by simple Drude model \cite{kang2009terahertz} and suggests a distinctive behavior in PtBi that warrants deeper exploration and investigation. Experimental evidence has demonstrated that the spin-charge conversion efficiency of PtBi is twice that of pure Pt \cite{hong2018giant}. However, due to the influence of other factors such as reduced impedance or interfacial spin-current transmission, the THz signal amplitude of PtBi is comparatively smaller than that of pure Pt. Here, our main focus was to perform a comparative analysis of THz emission from emitters fabricated with different materials. The investigation of emitters with distinct compositions of Pt$_{1-x}$Bi$_{x}$ will be addressed in a separate study.

\section{Conclusion}
To optimize spintronic THz emitters performance, we conducted a comparative study on SiO$_2$/CoFeB/NM heterostructures, where the CoFeB layer thickness varied from 0 to 5nm, and different heavy metals (Pt, W, Au) and alloys (Pt$_{\%92}$Bi$_{\%8}$ and Ag$_{\%90}$Bi$_{\%10}$) served as the NM layer. Our investigation revealed a critical threshold at 2nm thickness for the CoFeB layer, beyond which the emitted THz signal amplitude saturated. Furthermore, the heterostructure with 2nm CoFeB and 2nm Pt exhibited the highest THz signal amplitude. Surprisingly, despite the Pt$_{\%92}$Bi$_{\%8}$ (2nm) emitter exhibiting only half the THz amplitude compared to the Pt (2nm) emitter, it demonstrated a higher central THz frequency and the largest THz bandwidth among all the emitters investigated in our study. This study provide a solid foundation for future studies on different compositions of Pt$_{1-x}$Bi$_{x}$ alloy aimed at achieving even higher wideband and frequency capabilities in THz emitters. To gain a deeper understanding of the behavior exhibited by the different emitters, we conducted time-domain THz spectroscopy to analyze their conductivity and impedance characteristics. Our findings demonstrate that thicker NM layers exhibit higher electron concentration and lower mobility, resulting in lower impedance and subsequently lower THz amplitude. The anomalous behavior of PtBi, characterized by a negative imaginary part of conductivity according to our applied model, highlights the need for in-depth investigations and potential modifications of the model to better comprehend and explain the unique properties exhibited by this material. 

\subsection{Acknowledgement}

This project has received funding from the European Union’s Horizon 2020 research and innovation program under grant agreement No 899559 (SpinAge).

\subsection{Availability of data}
The data that supports the findings of this study are available within the article.

\bibliography{main.bib}
\clearpage
\end{document}